\documentclass[pra,twocolumn,superscriptaddress]{revtex4}
\usepackage{subfigure}
\usepackage{amsmath}
\usepackage{pgfplots}
\pgfplotsset{compat=1.5}
\usepgfplotslibrary{groupplots}

\usepackage{color}

\usepackage{xcolor}
\usepackage{bm}
\usepackage{multirow}
\usepackage{placeins}
\usepackage{soul}
\usepackage{color,xcolor}
\usepackage[colorlinks,linkcolor=red,anchorcolor=blue,citecolor=blue,urlcolor=blue]{hyperref}
\usepackage{cleveref}

\newcommand{\sgz}{\hat{\sigma}^z}
\newcommand{\sgp}{\hat{\sigma}^+}
\newcommand{\sgm}{\hat{\sigma}^-}
\newcommand{\aop}{\hat{a}}
\newcommand{\adop}{\hat{a}^{\dagger}}
\newcommand{\bop}{\hat{b}}
\newcommand{\bdop}{\hat{b}^{\dagger}}
\newcommand{\cop}{\hat{c}}
\newcommand{\cdop}{\hat{c}^{\dagger}}
\newcommand{\fop}{\hat{d}}
\newcommand{\fdop}{\hat{d}^{\dagger}}

\newcommand{\Hop}{\hat{H}}

\newcommand{\rhoop}{\hat{\rho}}
\newcommand{\im}{{\rm i}}
\newcommand{\fidelity}{\mathcal{F}}

\newcommand{\hc}{\rm H.c.}
\newcommand{\Nchain}{N_{\rm chain}}
\newcommand{\diag}{\rm diag}

\begin{document}

\title{Population transfer via a finite temperature state}
\author{Wei Huang}
\affiliation{Guangxi Key Laboratory of Optoelectronic Information Processing, Guilin University of Electronic Technology, Guilin 541004, China}

%\author{Muzaffar Lone}
%\affiliation{Department of Physics, University of Kashmir, Srinagar, India}

\author{Baohua Zhu}
\affiliation{School of material science and engineering, Guilin University of Electronic Technology, Guilin 541004, China}

\author{Wei Wu}
\affiliation{Department of Physics, College of Liberal Arts and Sciences, National University of Defense Technology, Changsha 410073, China}

\author{Shan Yin }
\email{syin@guet.edu.cn}
\affiliation{Guangxi Key Laboratory of Optoelectronic Information Processing, Guilin University of Electronic Technology, Guilin 541004, China}

\author{Wentao Zhang}
\email{zhangwentao@guet.edu.cn}
\affiliation{Guangxi Key Laboratory of Optoelectronic Information Processing, Guilin University of Electronic Technology, Guilin 541004, China}

\author{Chu Guo}
\email{guochu604b@gmail.com}
\affiliation{Henan Key Laboratory of Quantum Information and Cryptography, Zhengzhou, Henan 450000, China}

\begin{abstract}
We study quantum population transfer via a common intermediate state initially in thermal equilibrium with a finite temperature $T$, exhibiting a multi-level  Stimulated Raman adiabatic passage structure. We consider two situations for the common intermediate state, namely a discrete two-level spin and a bosonic continuum. In both cases we show that the finite temperature strongly affects the efficiency of the population transfer. We also show in the discrete case that strong coupling with the intermediate state, or a longer duration of the controlled pulse would suppress the effect of finite temperature. In the continuous case, we adapt the thermofield-based chain-mapping matrix product states algorithm to study the time evolution of the system plus the continuum under time-dependent controlled pulses, which shows a great potential to be used to solve open quantum system problems in quantum optics.
\end{abstract}

\date{\today}
\maketitle

\address{}

\vspace{8mm}

\section{Introduction}

Stimulated Raman adiabatic passage (STIRAP) is one of the most important technologies to implement complete population transfer from an initial state to a target state via a common intermediate state \cite{Vitanov2001,Vitanov2017,Shore2017}. In the standard implementation of STIRAP, two controlled laser pulses in Gaussian shapes, namely the $P$ pulse and $S$ pulse, are used to couple the initial state and the target state to the intermediate state respectively. When the two pulses are applied in a counter-intuitive order, that is, the $S$ pulse occurs before (but overlapping) the $P$ pulse, complete population transfer could be achieved with negligible excitation of the intermediate state. As a result this technique is very robust against the noises in the pulses as well as the dissipation in the intermediate state.

Due to robustness of STIRAP, there are many applications in different quantum systems to achieve completed population transfer from one quantum state to another, such as quantum optics \cite{Huang2017}, ion-trap system \cite{Moller2007}, superconducting qubits \cite{Kumar2016, Siewert2006}, cavity system \cite{Ye2003} and quantum dots system \cite{Hohenester2000}.
Interestingly, STIRAP technique can be employed not only in quantum systems but also in some specific classical systems, since the equations of motions governing these systems are analogous to the Schrodinger equation.
For example, we can employ STIRAP to waveguide coupler to achieve complete transfer of intensity of light from input waveguide to output waveguide \cite{Longhi2006}.
STIRAP can also be used in surface plasmon polaritons (SPPs) coupler excited by light on curved graphene sheets \cite{Huang2018}, the integrated terahertz device \cite{Huang20192}, and wireless energy transfer \cite{Rangelov2012}.

Since its initial proposition in a standard three-level configuration, the setup of STIRAP has been generalized in various directions. For example, Fractional STIRAP \cite{Sangouard2004}, Bright-State STIRAP \cite{Shore2013}, Straddle STIRAP \cite{Vitanov19981, Vitanov19982}, Two-State STIRAP \cite{Vitanov2017} and Composite-Pulse STIRAP \cite{Torosov2013}. These developments mainly focus on enhancing the robustness of STIRAP, or applying STIRAP in more general scenarios of multiple energy levels. 

In this paper, we study the setup of straddle STIRAP where population transfers from one energy level to another via multiple intermediate energy levels. It has been shown that complete population transfer could be achieved as long as the couplings between the two energy levels and the intermediate energy levels satisfy certain conditions~\cite{Vitanov1998,VitanovStenholm1999}. In~\cite{HuangGuo2020}, it is further shown that near-perfect population transfer could also be achieved for a finite-width continuum of intermediate states, and it is robust under moderate dissipation. However, to our best knowledge, most of the STIRAP-related works have assumed that the intermediate energy levels are initially in unoccupied (vacuum) states. In real applications, a frequently met situation is that the intermediate levels are initial in the thermal equilibrium state. For example, two spins coupled via an optical fiber, or an optical cavity\cite{Pulido2008} (or chain of cavities) initially in thermal equilibrium. In such cases, the excitations in the intermediate levels may participate in and intertwine the process, thus destroying the previous physical picture for STIRAP. Here we fill this gap by directly studying STIRAP-like population transfer via an intermediate thermal state. We mainly focus on two different setups: 1) Population transfer via a discrete two-level system initially in a thermal state with a temperature $T$ and 2) Population transfer via a bosonic thermal continuum. We study the effect of a finite-temperature intermediate state on the population transfer efficiency by numerically solving the quantum Liouville equation in those setups.

Our paper is organized as follows. In the Sec.\ref{sec:discretemodel}, we introduce the discrete version of the model which considers population transfer via a two-level system initially in a thermal state , and show the effect of the finite temperature on the efficiency of population transfer.  In Sec.\ref{sec:continuousmodel}, we introduce the continuous version of the model which considers population transfer via a thermal bosonic continuum and show the effect of the finite temperature in this case. We conclude in Sec.\ref{sec:summary}.

%  the two qubits coupling model via single spin or continuum bosonic bath with temperature and we apply the straddle STIRAP to this model.
% Subsequently, we demonstrate that the fidelity of population transfer between two qubits decreasing, with temperature increasing (as shown in Fig. 2).
% In the next section, we apply the continuum parallel STIRAP to the two qubits coupling with continuum common bosonic bath which has finite temperature to the common bath (see Fig. 3).
% We illustrate that thermal effect weaken the fidelity of population transfer in this model (see Fig. 4).
% In addition, we obtain the robustness of continuum parallel STIRAP in this model as shown in Fig. 5.

% Our paper is organized as follows. In Sec.\ref{sec:model}, we introduce our model and the equation of motion for the straddle-STIRAP via a continuum. In Sec.\ref{sec:result}, we numerically solve the quantum master equation for our model, and show the effectiveness of the popular transfer against changing the parameters of the model. We conclude in Sec.\ref{sec:summary}.

\begin{figure} [htbp]
\centering
\includegraphics[width=\columnwidth]{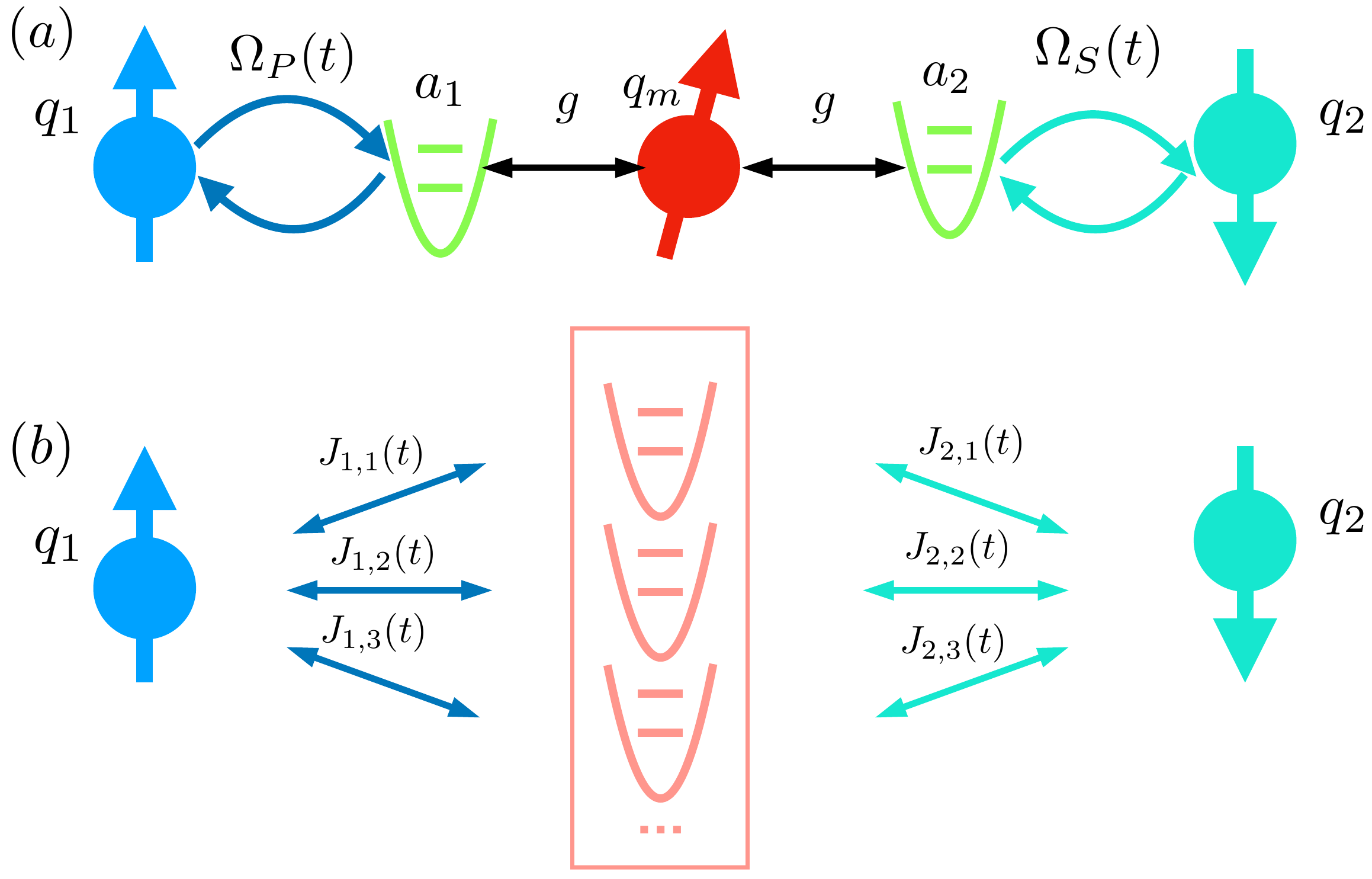}
\caption{(a) Population transfer between two qubits $q_1$ and $q_2$. The two qubits are coupled to two bosonic modes $a_1$ and $a_2$ via two controlled laser pulses $\Omega_P(t)$ and $\Omega_S(t)$ respectively. The two bosonic modes $a_1$ and $a_2$ are then coupled to a common intermediate spin $q_m$ with a strength $g$, which is initially in a thermal state with temperature $T$. (b) Population transfer between two qubits $q_1$ and $q_2$ which couple to a common intermediate bosonic continuum. The bosonic continuum is initially in a thermal distribution with temperature $T$.} \label{fig:fig1}
\end{figure}

\section{Discrete intermediate state}\label{sec:discretemodel}

First we consider population transfer via a discrete thermal state. For simplicity, we consider two spins which are coupled to two bosonic modes which act as the `flying qubits'. The two bosonic modes are then coupled to a common intermediate spin. The Hamiltonian of the whole system can be written as
\begin{align} \label{eq:hamiltonian}
\Hop(t) =& \frac{\omega_{q,1}}{2}  \sgz_1 + \frac{\omega_{q,2}}{2}  \sgz_2 + \omega_{a,1} \adop_1\aop_1 + \omega_{a,2}\adop_2\aop_2 + \nonumber \\
& \Omega_P(t) (\adop_1 \sgm_{1} + \aop_1 \sgp_{1}) + \Omega_S(t) (\adop_2 \sgm_{2} + \aop_2 \sgp_{2}) + \nonumber \\
&\frac{\omega_m}{2} \sgz_m + g \left(\adop_1\sgm_m + \aop_1\sgp_m\right) + g \left(\adop_2\sgm_m + \aop_2\sgp_m\right),
\end{align}
where $\omega_{q,1}$ and $\omega_{q,2}$ are the energy differences of the two qubits, $\omega_{a, 1}$ and $\omega_{a,2}$ are the frequencies of the two bosonic modes, and the time-dependent couplings $\Omega_P(t)$ and $\Omega_S(t)$ between the two qubits and the bosonic mode are induced by two controlled pulses, which are defined as
\begin{align}
\Omega_P(t) &= \Omega\exp\left(\frac{-\left(t - \tau/2\right)^2}{\tau_0^2}\right); \\
\Omega_S(t) &= \Omega\exp\left(\frac{-\left(t + \tau/2\right)^2}{\tau_0^2}\right),
\end{align}
with $\tau_0$ the standard deviation of Gaussian pulses, $\tau$ the time delay between the two pulses, and $\Omega$ the maximum strength of the pulses. $\omega_m$ is the energy difference of the intermediate spin, and $g$ is the coupling strength between the intermediate spin and the two bosonic modes. We have set $\hbar=1$. The dynamics of this system is described by the quantum Liouville equation
\begin{align}\label{eq:liouville}
\frac{d\rhoop(t)}{dt} = -\im [\Hop(t), \rhoop].
\end{align}
In the rest of this work we will always use the resonant condition such that $\omega_{q, 1}=\omega_{q, 2}=\omega_{a, 1}=\omega_{a, 2}=\omega_m$. The bosonic modes are not occupied initially while the intermediate spin is assumed to be in a thermal state with a temperature $T$, that is
\begin{align}
\rhoop^m = \frac{1}{1+e^{-\beta \omega_m}} \vert 0^m\rangle\langle 0^m\vert + \frac{e^{-\beta \omega_m}}{1+e^{-\beta \omega_m}} \vert 1^m\rangle\langle 1^m\vert.
\end{align}
Here $\beta$ is the inverse temperature $\beta=1/T$ and we have set the Boltzmann constant $k_B=1$. Thus the initial state of the whole system can be written as
\begin{align}
\rhoop_i = \vert 1^{q_1} \rangle\langle 1^{q_1}\vert \otimes \vert 0^{a_1} \rangle\langle 0^{a_1}\vert \otimes \rhoop^m \otimes \vert 0^{a_2} \rangle\langle 0^{a_2}\vert \otimes \vert 0^{q_2} \rangle\langle 0^{q_2}\vert,
\end{align}
where $\vert 0^{q_i}\rangle$ ($\vert 1^{q_i}\rangle$) means the ground (excited) state of the spin $q_i$, and $\vert 0^{a_i} \rangle$ means the vacuum state for the bosonic mode $a_i$, with $i=1,2$. The final state after the time evolution is denoted as $\rhoop_f$, namely $\rhoop_f = \rhoop(\infty)$. Moreover, we denote $\fidelity_i(t)$ as the occupation on the excited state of the spin $q_i$, that is,
\begin{align}
\fidelity_1(t) &= \langle 1^{q_1} \vert \rhoop^{q_1}(t) \vert  1^{q_1} \rangle, \\
\fidelity_2(t) &= \langle 1^{q_2} \vert \rhoop^{q_2}(t) \vert  1^{q_2} \rangle,
\end{align}
where $\rhoop^{q_i}(t)$ means the reduced density operator of the spin $q_i$. In our setup, we have $\fidelity_1(-\infty)=1$ and $\fidelity_2(-\infty)=0$, and perfect population is achieved if $\fidelity_1(-\infty)=0$ and $\fidelity_2(-\infty)=1$. In the following we will use $\fidelity = \fidelity_2(\infty)$ to denote the final fidelity.

\begin{figure}%[H]
\includegraphics[width=\columnwidth]{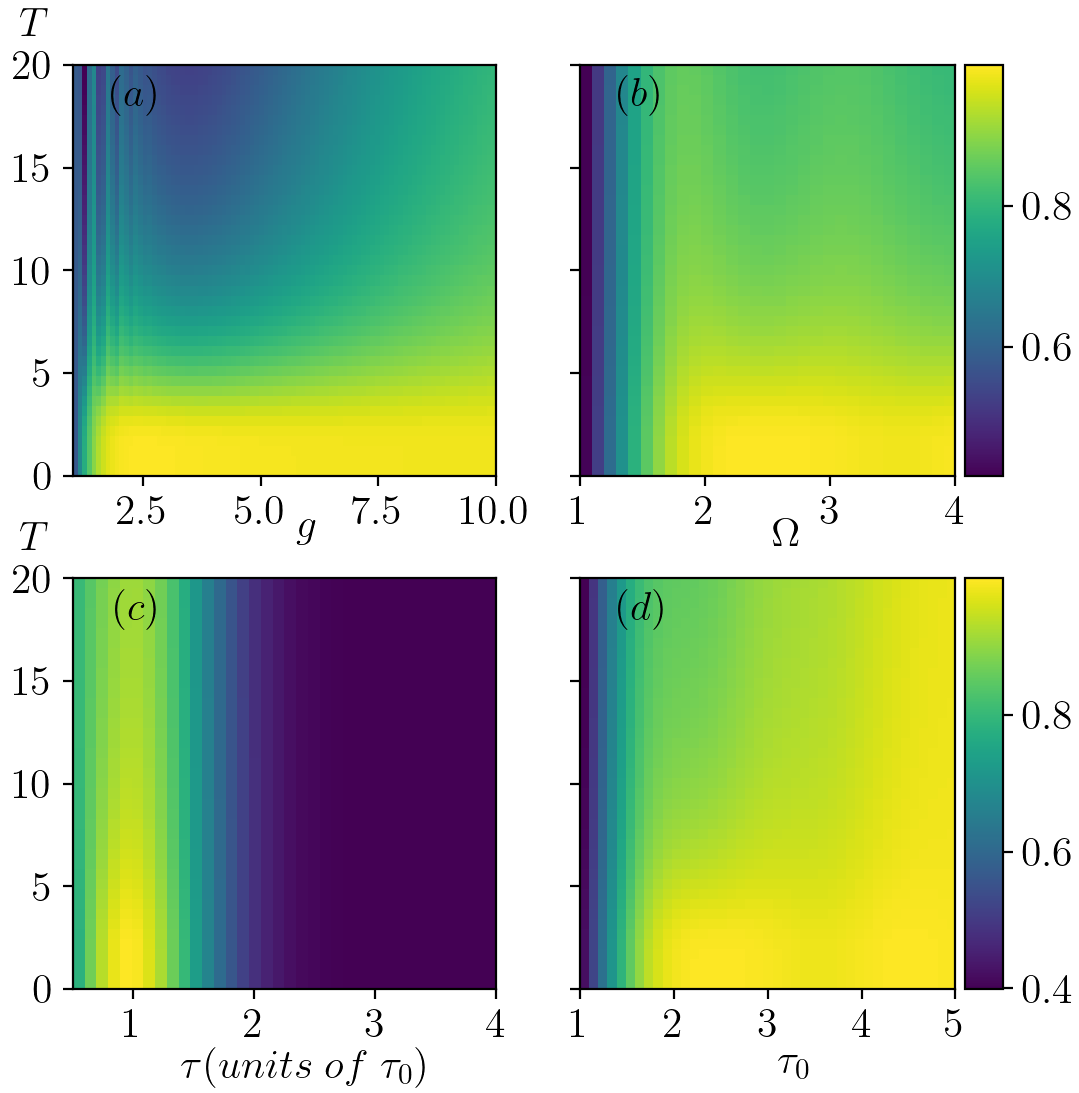}
\caption{Dependence of $\fidelity$ on temperature $T$ in the discrete case, where $T$ ranges from $0$ to $20$ in all panels. (a) $\fidelity$ as a function of $g$ and $T$, where $g$ ranges from $1$ to $10$. (b) $\fidelity$ as a function of $\Omega$ and $T$, where $\Omega$ ranges from $1$ to $4$. (c) $\fidelity$ as a function of $\tau$ and $T$, where $\tau$ ranges from $0.5\tau_0$ to $4\tau_0$. (d) $\fidelity$ as a function of $\tau_0$ and $T$, where $\tau_0$ ranges from $1$ to $5$. The other parameters used in all panels, unless otherwise specified, are $g=10$, $\Omega=2$, $\tau=1$ and $\tau_0=2$.}
   \label{fig:fig2}
\end{figure}

To show the effect of the temperature $T$ and the interplay between $T$ and the other parameters, we simulate the dynamics of Eq.~\ref{eq:liouville} in a wide parameter range, and the results are shown in Fig.~\ref{fig:fig2}. In Fig.\ref{fig:fig2}(a), we show the dependence of the final fidelity $\fidelity$ on the temperature $T$ and the coupling strength $g$ between the bosonic modes and the intermediate spin. We can see that $\fidelity$ is greatly suppressed when increasing $T$, showing that a highly occupied excited state would strongly affect the efficiency of population transfer. We can also see that $\fidelity$ slightly goes up with $g$, especially at higher temperature. This is expected since a standard requirement for perfect STIRAP is the strong coupling between the initial (final) states with the intermediate states. This result is interesting in that it show that although STIRAP is known to be robust against the dissipation of the intermediate state, however it will be strongly affected if the intermediate spin is in a highly mixed state. In Fig.~\ref{fig:fig2}(b), we show the dependence of $\fidelity$ on $T$ and the maximum amplitude of the laser pulse $\Omega$, with $g=10, \tau=1, \tau_0=2$. We see a similar effect to Fig.~\ref{fig:fig2}(a) that $\fidelity$ increases with larger $\Omega$ and decreases with larger $T$. This is because $\Omega$ play a similar role as $g$ which determines the coupling strength between the initial (final) state and the intermediate spin. In Fig.~\ref{fig:fig2}(c), we show the dependence of $\fidelity$ on $T$ and the time delay $\tau$. We can see that there is a pick around $\tau=\tau_0$, and since the coupling strength $g=10$ and $\Omega=2$ are large enough, relatively high population transfer efficiency could still be achieved at high temperature. In Fig.~\ref{fig:fig2}(d), we show the dependence of $\fidelity$ on $T$ and the period of driving $\tau_0$. We can see that $\fidelity$ is larger with larger $\tau_0$. This is expected since larger $\tau_0$ means the time evolution is slower, thus more adiabatic, which is another standard requirement of STIRAP. For large $T$, population transfer efficiency is slightly suppressed but much less significant than in cases of Fig.\ref{fig:fig2}(a, b). 

\section{Continuous intermediate state}\label{sec:continuousmodel}

Now we further consider the case that the two qubits are coupled via an intermediate finite-temperature bosonic continuum. The Hamiltonian of the whole system can be written as
\begin{align}\label{eq:continuousH}
\Hop(t) = &\frac{\omega_{q,1}}{2}  \sgz_1 + \frac{\omega_{q,2}}{2}  \sgz_2 + \int d\omega \omega \bdop_{\omega}\bop_{\omega} +  \nonumber \\
& \Omega_P(t) \int d\omega \sqrt{\mathcal{J}(\omega)}\left(\sgp_1 \bop_{\omega} + \sgm_1\bdop_{\omega} \right) + \nonumber \\
& \Omega_S(t) \int d\omega \sqrt{\mathcal{J}(\omega)}\left(\sgp_2 \bop_{\omega} + \sgm_2\bdop_{\omega} \right),
\end{align}
where $\mathcal{J}(\omega)$ is the spectrum function. We choose a simple sub-ohmic spectrum as
\begin{align}
\mathcal{J}(\omega) = \sqrt{\omega},
\end{align}
and we also choose a sharp cut-off $\omega_c$ such that $\mathcal{J}(\omega)=0$ for $\omega > \omega_c$, as a signature of a finite-width continuum. In comparison with the discrete case considered in Sec.\ref{sec:discretemodel}, we have removed the two intermediate `flying qubits' $a_1$ and $a_2$, which will allow an easier numeric treatment while the resulting physics is still similar. In case the continuum is initially in the zero temperature state, the dynamics of Eq.~\ref{eq:continuousH} can be easily solved based on a discretization of the continuum and an exact diagonalization approach since only the single excitation sector needs to be considered~\cite{HuangGuo2020}. However, for a finite temperature $T$, the continuum is a mixture of different bosonic particles and the Hilbert space size is in general exponentially large. As a result exact diagonalization would be impossible in this case. Moreover, in this case a Markovian quantum master equation, such as the Lindblad equation~\cite{Lindblad1976,Gorini1976}, would likely be problematic since here we consider strong system-continuum coupling.

In recent years, there is a growing activity to use the system-bath approach in combination with matrix product states method to study the dynamics of open quantum systems. The system and the bath are evolved together as a whole, and the dynamics of the system is obtained by tracing out the bath degrees of freedoms. Here we use a thermofield-based chain-mapping matrix product states algorithm (TCMPS)~\cite{InesBanuls2015,MascarenhasVega2017, GuoPoletti2018, FuggerArrigoni2018,SchwarzWeichselbaum2018,XuPoletti2019,ChenPoletti2020} to study the dynamics of the system plus bath which the the continuum in our case. The main advantage of this method is that the finite temperature bath is mapped into another enlarged bath which is initially at zero temperature, thus in favoring of a MPS simulation. TCMPS include three major steps: 1) Discretization of the bath~\cite{InesWolf2015} , for which we use a simple linear discretization scheme with a frequency step size $\delta$, the discretized Hamiltonian after this step would be
\begin{align}
\Hop^{\rm dis}(t) =& \frac{\omega_{q,1}}{2}  \sgz_1 + \frac{\omega_{q,2}}{2}  \sgz_2 + \sum_{j=1}^N \omega_j \bdop_j\bop_j \nonumber \\
&+ \Omega_P(t) \sum_{j=1}^N J_{j}\left(\aop_1\bdop_j + \adop_1 \bop_j \right) \nonumber \\
&+ \Omega_S(t) \sum_{j=1}^N J_{j}\left(\aop_2\bdop_j + \adop_2 \bop_j \right),
\end{align}
where we have used $N = \omega_c/\delta$, $\omega_j = j\delta$, $\bop_j = \bop(\omega_j)$, $\bdop_j = \bdop(\omega_j)$, $J_{j} = \sqrt{\mathcal{J}(\omega_j)\delta}$. The time-dependent couplings $J_{1,j}(t)$ and $J_{2,j}(t)$ in Fig.~\ref{fig:fig1}(b) correspond to $\Omega_P(t)J_j$ and $\Omega_S(t)J_j$ respectively. In the limit $N\rightarrow \infty$, $\Hop^{\rm dis}(t)$ is equivalent to $\Hop(t)$~\cite{BullaPruschke2008,InesWolf2015}; 2) Thermofield transformation maps the bath of $N$ bosonic modes into an enlarged but equivalent bath with $2N$ bosonic modes, and at the same time the thermal state corresponding to the original bath is mapped into the vacuum state of the enlarged bath. The Hamiltonian after this step is
\begin{align}
\Hop^{\rm T}(t) =& \frac{\omega_{q,1}}{2}  \sgz_1 + \frac{\omega_{q,2}}{2}  \sgz_2 + \sum_{j=1}^N \omega_j \left(\cdop_{1,j}\cop_{1,j} - \cdop_{2,j}\cop_{2,j}\right) \nonumber \\
&+\Omega_P(t)\sum_{j=1}^N g_{1,j}\left(\aop_1\cdop_{1,j} + \adop_1\cop_{1,j} \right) \nonumber  \\
&+\Omega_P(t)\sum_{j=1}^N g_{2,j}\left(\aop_1\cop_{2,j} + \adop_1\cdop_{2,j} \right) \nonumber  \\
&+\Omega_S(t)\sum_{j=1}^N g_{1,j}\left(\aop_2\cdop_{1,j} + \adop_2\cop_{1,j} \right) \nonumber  \\
&+\Omega_S(t)\sum_{j=1}^N g_{2,j}\left(\aop_2\cop_{2,j} + \adop_2\cdop_{2,j} \right),
\end{align}
where $g_{1,j} = J_j \cosh(\theta_j)$ and $g_{2,j} = J_j\sinh(\theta_j)$, with $\cosh(\theta_j)=\sqrt{1+n(\omega_j)}$, $\sinh(\theta_j) = \sqrt{n(\omega_j)}$ and $n(\omega) = 1/\left(e^{\beta\omega}-1\right)$ to be the Bose-Einstein distribution.
3) Star to chain mapping, which maps the system-bath from the star configuration into a chain configuration. The final  Hamiltonian after those three steps would be
\begin{align}
\Hop^{\rm TC}(t) = & \frac{\omega_{q,1}}{2}  \sgz_1 + \frac{\omega_{q,2}}{2}  \sgz_2 + \sum_{\nu=1}^2 \sum_{j=1}^{\Nchain} \alpha_{\nu, j} \fdop_{\nu, j}\fop_{\nu, j} \nonumber \\
&+ \Omega_P(t) \sum_{j=1}^{\Nchain}\left(\beta_{1, 1}\aop_1\fdop_{1, j} + \beta_{2, 1}\aop_1\fop_{2, j} + \hc \right) \nonumber \\
&+ \Omega_S(t) \sum_{j=1}^{\Nchain}\left(\beta_{1, 1}\aop_2\fdop_{1, j} + \beta_{2, 1}\aop_2\fop_{2, j} + \hc \right) \nonumber \\
&+ \sum_{\nu=1}^2\sum_{j=1}^{\Nchain-1}\beta_{\nu, j+1}\left(\fdop_{\nu, j}\fop_{\nu, j+1} + \hc \right),
\end{align}
where $\alpha_{1, j}$ and $\beta_{1,j}$ are the diagonal terms and off-diagonal terms resulting from the Lanczos tri-diagonalization of the diagonal matrix $\diag([\omega_1, \omega_2, \dots, \omega_N])$ with the initial vector $[g_{1,1}, g_{1,2}, \dots, g_{1, N}]$, while $\alpha_{2, j}$ and $\beta_{2,j}$ are the diagonal terms and off-diagonal terms resulting from the Lanczos tri-diagonalization of the diagonal matrix $\diag([-\omega_1, -\omega_2, \dots, -\omega_N])$ with the initial vector $[g_{2,1}, g_{2,2}, \dots, g_{2, N}]$~\cite{GuoPoletti2018}. The size of the vectors $\alpha_{2,j}$ and $\beta_{2,j}$, denoted as $\Nchain$, is usually chosen to be less than $N$. To the best of our knowledge, this work is the first time to apply TCMPS to study an open quantum system with time-dependent driving.

\begin{figure}%[H]
\includegraphics[width=\columnwidth]{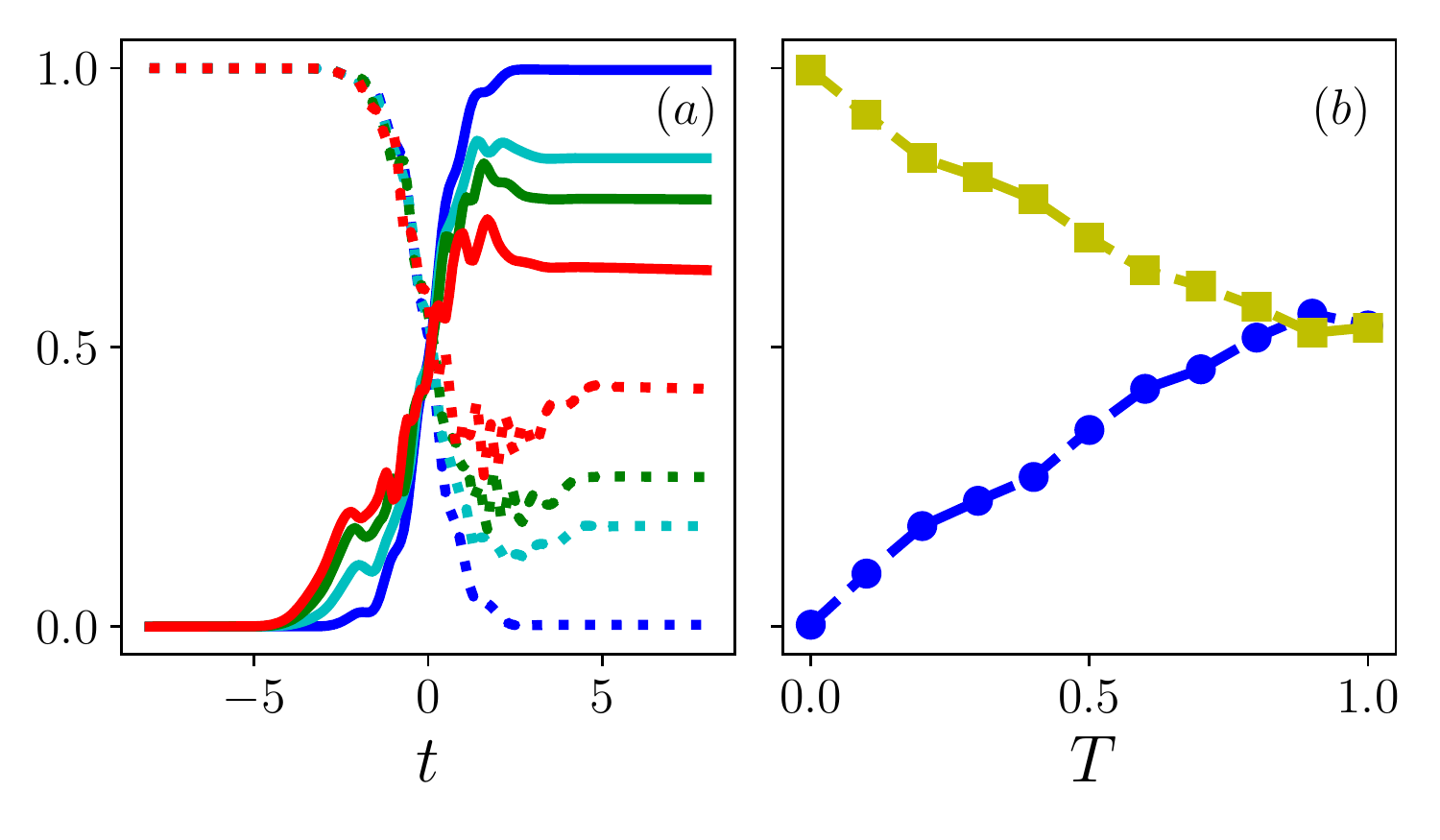}
\caption{Population transfer via a thermal bosonic continuum. (a) The solid lines from top down correspond to $T=0, 0.2, 0.4, 0.6$ respectively, which plot $\fidelity_1$ as a function of time $t$. The dotted lines from down to top correspond to $T=0, 0.2, 0.4, 0.6$ respectively, which plot $\fidelity_2$ as a function of time $t$. (b) The blue dashed line with circle corresponds to $\fidelity_1(\infty)$ as a function of temperature $T$, while the yellow dashed line with square corresponds to $\fidelity_2(\infty)$ as a function of temperature $T$. }
   \label{fig:fig3}
\end{figure}

We then evolve $\Hop^{\rm TC}(t)$ with the same initial state for the two spins as for the discrete case, and vacuum state for the enlarged continuum corresponding to the set of modes $\fop_{\nu, j}$. In our simulations we have chosen $\omega_c=2$, $\delta=0.01$, $\Nchain=50$, a time step size $dt=0.01$ and we have kept $400$ auxiliary states. The largest singular value truncation error observed during the time evolution is of the order $10^{-4}$. The simulation results are shown in Fig.~\ref{fig:fig3}. In Fig.~\ref{fig:fig3}(a), we plot $\fidelity_1(t)$ and $\fidelity_2(t)$ as a function of time $t$, we can see that in case of $T=0$, almost perfect population transfer can be achieved, which is also shown in~\cite{HuangGuo2020}. As $T$ increases, the efficiency of population transfer goes down significantly. In Fig.~\ref{fig:fig3}(b), we plot $\fidelity_1(\infty)$ and $\fidelity_2(\infty)$ as a function of the temperature $T$, from which we can see more clearly that the efficiently of population transfer goes down significantly when $T$ increases. At $T=1$, only about half of the population are successfully transferred from $q_1$ to $q_2$. These results show that for STIRAP via an infinite number of intermediate states, the non-zero temperature strongly affects the population transfer efficiency. 

\section{conclusion} \label{sec:summary}
We propose two models to study quantum population transfer between two spins via an intermediate state which is initially in thermal equilibrium. In the first case, we consider a discrete model where the two spins are coupled to two bosonic modes by two controlled pulses $\Omega_P(t)$ and $\Omega_S(t)$ which act as `flying qubits', which are then coupled to a common intermediate spin initially in a thermal state. In the second case, we consider a continuous model where the two spins are directly coupled to a thermal bosonic continuum by the two controlled pulses. In both cases, we show that the efficiency of the population transfer is strongly dependent on the finite temperature of the intermediate state, in contrast with previous results that the population transfer efficiency is robust against the details of the intermediate states as long as certain control parameters are well tuned.

Moreover, in this work we have adapted the TCMPS method, which is a recently developed numeric technique used to solve open quantum many-body systems, to study quantum population transfer via a thermal bosonic continuum. Our results show that TCMPS could be a perfect numerical tool to study open quantum optics problems in presence of a finite temperature environment and time-dependent driving.

% We propose a model to study population transfer where the intermediate states is a bosonic continuum. The model consists of two spins which are coupled to two bosonic modes with a dynamical coupling strength $\Omega_P(t)$ and $\Omega_S(t)$, and the two bosonic modes are indirectly coupled through a bosonic continuum. We show the effects on the efficiency of population tranfer when tuning the the coupling strength between the bosonic modes with the continuum, as well as the various control parameters of the laser pulses. We also consider the case that when the continuum subject to a constant particle loss rate, and show that efficient population transfer can still be achieved with a moderate dissipation. We believe that this finding will be improve the high efficient transfer information in quantum information processing in future.

\section{Acknowledgement}
This work is acknowledged for funding National Science and Technology Major Project (grant no. 2017ZX02101007-003); National Natural Science Foundation of China (grant no. 61565004; 6166500; 61965005); the Natural Science Foundation of Guangxi Province (Nos. 2017GXNSFBA198116 and 2018GXNSFAA281163); the Science and Technology Program of Guangxi Province (No. 2018AD19058). W.H. is acknowledged for funding from Guangxi oversea 100 talent project and W.Z. is acknowledged for funding from Guangxi distinguished expert project. C. G acknowledges support from National Natural Science Foundation of China under Grants No. 11805279.

\bibliographystyle{apsrev4-1}
\bibliography{refs}

\end{document}